\documentclass[sigconf]{acmart}
\usepackage{bookmark}
\usepackage{subfig}
\AtBeginDocument{%
  \providecommand\BibTeX{{%
    \normalfont B\kern-0.5em{\scshape i\kern-0.25em b}\kern-0.8em\TeX}}}

\copyrightyear{2022}
\acmYear{2022}
\setcopyright{rightsretained}

\acmConference[AIES'22]{Proceedings of the 2022 AAAI/ACM Conference on AI, Ethics, and Society}{August 1--3, 2022}{Oxford, United Kingdom}
\acmBooktitle{Proceedings of the 2022 AAAI/ACM Conference on AI, Ethics, and Society (AIES'22), August 1--3, 2022, Oxford, United Kingdom}\acmDOI{10.1145/3514094.3534166}
\acmISBN{978-1-4503-9247-1/22/08}

\usepackage{etoolbox}
\makeatletter
\patchcmd{\maketitle}{\@copyrightpermission}{
   \begin{minipage}{0.3\columnwidth}
     \href{https://creativecommons.org/licenses/by/4.0/}{\includegraphics[width=0.90\textwidth]{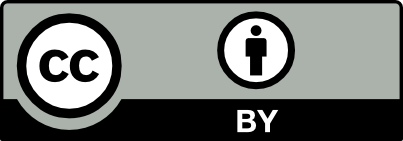}}
   \end{minipage}\hfill
   \begin{minipage}{0.7\columnwidth}
     \href{https://creativecommons.org/licenses/by/4.0/}{This work is licensed under a Creative Commons Attribution International 4.0 License.}
   \end{minipage}
 
   \vspace{5pt}
}{}{}

\makeatother

\title{How Algorithms Shape the Distribution of Political Advertising: Case Studies of Facebook, Google, and TikTok}

\author{Orestis Papakyriakopoulos}
\affiliation{%
  \institution{Center for Information Technology Policy}
  \country{Princeton University}
}
\email{orestis@princeton.edu}

\author{Christelle Tessono}
\affiliation{%
  \institution{Center for Information Technology Policy}
    \country{Princeton University}
}
\email{ct8474@princeton.edu}

\author{Arvind Narayanan}
\affiliation{%
  \institution{Center for Information Technology Policy}
    \country{Princeton University}
}
\email{arvindn@cs.princeton.edu}

\author{Mihir Kshirsagar}
\affiliation{%
  \institution{Center for Information Technology Policy
  }
  \country{Princeton University}
}
\email{mihir@princeton.edu}

\renewcommand{\shortauthors}{Papakyriakopoulos et al.}

\begin{abstract}
Online platforms play an increasingly important role in shaping democracy by influencing the distribution of political information to the electorate. In recent years, political campaigns have spent heavily on the platforms’ algorithmic tools to target voters with online advertising. While the public interest in understanding how platforms perform the task of shaping the political discourse has never been higher, the efforts of the major platforms to make the necessary disclosures to understand their practices falls woefully short. In this study, we collect and analyze a dataset containing over 800,000 ads and 2.5 million videos about the 2020 U.S. presidential election from Facebook, Google, and TikTok. We conduct the first large scale data analysis of public data to critically evaluate how these platforms amplified or moderated the distribution of political advertisements.\footnote{Our dataset and source code is available for public use under http://campaigndisclosures.princeton.edu/} We conclude with recommendations for how to improve the disclosures so that the public can hold the platforms and political advertisers accountable. 

\end{abstract}



\begin{document}

\maketitle

\section{Introduction}

As online advertising becomes a crucial  part of political campaigns \cite{hersh2015hacking,kreiss2016prototype}, the platforms’ control over their communication infrastructure makes them key political actors \cite{plantin2018infrastructure} and gives them a power over the political discourse that goes beyond what the traditional definition of ``platform'' might denote \cite{gillespie2010politics}. Importantly, these platforms are not neutral carriers of political ads, but play a more active role in amplifying or moderating the reach of those political messages. But the platforms have not disclosed data that would allow for meaningful public oversight of their actions. In 2018, in an attempt to stave off regulation, some platforms begun to voluntarily create libraries of political advertisements and moderation decisions \cite{fowler2020online}. 

In this study, we evaluate the voluntary disclosures \cite{google_ads,facebook_political_ads} made by online platforms to understand how the platforms influence the distribution of the political ads. We conduct the first large scale data analysis of political ads in the 2020 U.S. presidential elections to investigate the practices of three platforms - Facebook, Google, and TikTok.

\begin{figure}[!h]
    \centering
    \includegraphics[width=0.47\textwidth]{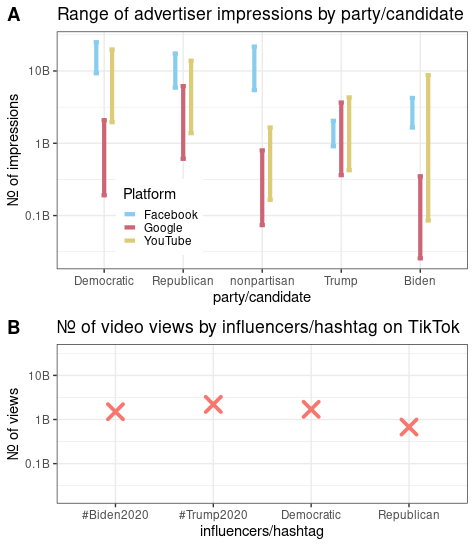}
    \caption{ Overview of political content reach by platform in our dataset, for the two months up to the election day. On the top, Figure A depicts Facebook in blue, YouTube in black, and Google in yellow. Facebook \& Google do not disclose the exact number of impressions, but a range within which the value falls in.  On the bottom, Figure B depicts the number of views of videos containing the hashtags \#Biden2020 and \#Trump2020 on TikTok in our dataset. It also presents the number of views for videos produced by 96 political influencers (see Table \ref{influencers}.}
    \label{overview}
\end{figure}

One clear indication of the importance of online platforms to political campaigns is to see how campaigns have shifted their spending to online advertising. In 2008, the first significant digital campaign spent roughly \$20 million on online advertising, which amounted to 0.4\% of the total money spent on campaigning \cite{erickson_2016}. In the 2020 election cycle, the campaigns spent more than \$2.8 billion, or 20\% of the campaign budget \cite{statista} on the major platforms. As shown in figure \ref{overview}, this spending generated billions of impressions for political ads placed in the two months prior to the US 2020 elections. 

Mindful of the risk of elevating transparency to become the supreme value in democratic politics \cite{schudson2020shortcomings}, we do not focus on transparency for its own sake. Instead, we develop a framework that identifies the desirable properties for these disclosures that serve broader democratic values, and measuring the effectiveness of the current platform disclosures against those properties. Two straightforward research questions lie at the heart of our inquiry:

\begin{description}

\item[\textbf{RQ1:}]What do the ad libraries tell us about how the algorithmic tools were used to distribute political ads?

\item[\textbf{RQ2:}]Can we interpret how platforms applied their moderation policies to political ads?

\end{description}

\subsection{Contributions}

\begin{itemize}


\item We develop and apply a three part evaluative framework for measuring the effectiveness of platform disclosures (section \ref{evaluative}).

\item We attempt to reverse-engineer platforms' ad targeting tools using the available data to assess how they influence the distribution of content (section \ref{distribution}).  


\item Our statistical analysis suggests that the platforms charge different campaigns different rates for targeting specific regions and demographics (section \ref{distribution2}). 

\item As a whole, we demonstrate how the data provided by the platforms is noisy and incomplete so it is difficult to draw definitive conclusions about how they operate (section \ref{distribution3}).


\item There are pervasive inconsistencies in how the platforms implement their ad moderation policies. We detected numerous instances where ads are moderated after already having generated millions of impressions, or where some ads are flagged for moderation when others with the same content are not (section \ref{moderation}).  

\end{itemize}

\section{Related Work}

Online platforms such as Facebook, Google, and TikTok play an increasingly prominent role in shaping political discourse \cite{kreiss2018technology}. One widely studied aspect is the impact of the platform practices on user behavior \cite{corbyn2012facebook,persily2020social,susser2021measuring}. Another is how political actors utilize platforms in their political campaigns \cite{kreiss2016seizing}. But the role of platforms' algorithmic tools as intermediaries that shape the political discourse has not been studied as extensively as there is limited visibility into their practices \cite{edelson2019analysis,ghosh2019analyzing}.

We build on the works from Kreiss, Mcgregor \& Barrett \cite{kreiss2018technology,kreiss2019arbiters,kreiss2020democratic}, exploring the role of online platforms in shaping political communication, and the works of Fowler et al.  \cite{fowler2018political, fowler_2021,fowler2021political} and West\cite{west2017air}, who provide historical information about political advertising and regulatory changes. We also draw on work by researchers who describe the content of ad libraries (E.g. Edelson et al. \cite{edelson2019analysis} for the US, Dubois et al for Canada \cite{dubois2021micro} and Medina et al. for Germany \cite{medina2020exploring}). Like \cite{andreou2018investigating,leerssen2021news,silva2020facebook, le2022audit}, we assess the data quality of the libraries to uncover misleading, incomplete, or wrong information. 

Notably, our work is different from and complementary to the attempts to understand the role of the platforms in democracy through direct agreements between researchers and platforms \cite{guess2021cracking} that draw on the detailed user data available to the platform. Our study is deliberately limited to the data that the platforms make publicly accessible.


As we demonstrate in our analysis below, the platforms use complex, and opaque algorithms to price and distribute ads (e.g \cite{toubiana2010adnostic,matias2021software}). We build on prior research studies that show how these algorithmic tools can result in discriminatory and biased outcomes \cite{ali2021measuring,ali2021ad,ali2019discrimination}, and can have differential effects on the user population \cite{chen2015making,kruikemeier2016political}. 


Finally, our recommendations for appropriate disclosures rely on design frameworks (e.g. \cite{finocchiaro2021bridging,benthall2021artificial,simons2020utilities}) that
seek to give users the ability to evaluate and understand how algorithmic tools influence them \cite{milano2021epistemic,neff2020bad,vecchione2021algorithmic}.

\section{Evaluative Criteria} \label{evaluative}

Government regulations for online political advertising have stalled in the United States \cite{persily2020social}. As a result, we do not have legal standards with which to evaluate the current disclosures by the platforms. Nevertheless, we extract three potential criteria to measure the effectiveness of the disclosures:

\begin{itemize}
    \item First, do the disclosures meet the platforms' self-described objective of making political advertisers accountable? 
    
    \item Second, how do the platforms' disclosures compare against what the law requires for radio and television broadcasters? 
    
    \item Third, do the platforms disclose all that they know about the ad targeting criteria, the audience for the ads, and how their algorithms distribute or moderate content?   
\end{itemize}

\subsection{Self-Imposed Standards}
In 2018, facing potential regulation such as the proposed Honest Ads Act \cite{congress}, several online platforms chose to create ad libraries of political campaign materials. For Facebook, Google, and YouTube, these libraries provide some basic information about who placed ads, their content, how they were distributed, and whether they were moderated (table \ref{ad_policies}). TikTok recently created an ads library \cite{tiktok_ads_library}, but the company disavows carrying ads about political issues and it does not disclose how it moderates political content.  

\begin{table}[]
\caption{Platform specific strategies in distributing and moderating political content, showing how each platform defines political content, the access to targeting tools, the presence of an ad library, and their moderation practices. YouTube had the same ad policy as google, so we did not include a seperate entry for it. }
\resizebox{0.48\textwidth}{!}{\begin{tabular}{l|lll}
                               & \begin{tabular}[x]{@{}l@{}} \textbf{Facebook } \\ \end{tabular} & \begin{tabular}[x]{@{}l@{}} \textbf{Google } \\  \end{tabular} & \begin{tabular}[x]{@{}l@{}}  \textbf{TikTok }  \\ \end{tabular} \\ \hline
\textbf{Definition}            & Actor, Subject           & Actor, Issue                          & \begin{tabular}[x]{@{}l@{}}Actor, Subject,\\ Issue \end{tabular}\\[10pt]
\textbf{Targeting}             & Full                     & Restricted                     & None                     \\[10pt]
\begin{tabular}[x]{@{}l@{}}\textbf{Election Ad}\\ \textbf{libraries}\end{tabular}  &  \begin{tabular}[x]{@{}l@{}}ad cost,\\ impressions,\\ audience \\characteristics \\ (gender, age,\\ state) \end{tabular}                  &  \begin{tabular}[x]{@{}l@{}}ad cost,\\ impressions, \\targeting\\ parameters \\(gender,\\ age, location) \end{tabular}                                & \begin{tabular}[x]{@{}l@{}}No political \\content\end{tabular}                     \\[40pt]
\textbf{Moderation}            & Removal/ Label           & Removal/Label                  & Label/Algorithm         \\ \hline
\end{tabular}}
\label{ad_policies}
\end{table}

\textbf{Facebook}. Facebook’s Political Ad Policy restricts the ability to run electoral and issue ads to authorized advertisers only. Facebook states that the purpose of the ad library is to provide “advertising transparency by offering a comprehensive, searchable collection of all ads currently running from across Facebook apps and services, including Instagram.” Notably, as the policy explains, the advertising transparency is directed at “making political ads more transparent and advertisers more accountable” and not to hold the platform accountable for how it distributes or moderates the political content.

For ad moderation Facebook applies its general Advertising Policies and Community Standards. The Political Ads Policy falls under the Restricted Content section and consists of two policies: \textit{9.a Ads About Social Issues, Elections or Politics}, and \textit{9.b Disclaimers for Ads About Social Issues, Elections or Politics}. Article 9.a outlines that advertisers are required to complete Facebook’s authorization process, and failure to meet the reporting requirements may lead to restrictions such as the disabling of existing ads.  

\textbf{Google}. Google also launched its political Ad Library during summer 2018 and requires advertisers to be verified to publish ads. Like Facebook, the library’s purpose is vaguely described as providing “greater transparency in political advertising,” without disclosing what the transparency is being compared against or who is the subject of the transparency goals. It is clear that the subject of transparency is the political campaign and not the platform. The Google ads archive, which includes ads placed on the Google network (search engine, third party websites that use google ad tools, and other google services) and on YouTube, shows the content of each instance of the political ad, the advertiser, its cost and related impressions. It also shows which user groups in terms of age, gender, and location (up the zip code), were targeted by the advertisers. 

Google and YouTube's ad moderation policies are set forth in the platform’s Advertising Policies. Google has a specific category on Political Content, which is listed under the Restricted Content and Features section of their policy. In case that an ad gets removed by the platform, the content of it is replaced by a red banner in the archive, stating that the ad violated the platform’s terms \& conditions.

\textbf{TikTok}. TikTok does not have political ads in its library because it does not allow such ads on the platform. It explains that the ``the nature of paid political ads is not something we believe fits the TikTok platform experience.'' Nevertheless, from the content sample we analyze in this study (see section \ref{tiktok_methods}) we document a significant amount of political content shared on the platform around the elections. We observe that a lot of that content is generated by a group of influencers, some of which directly linked to political organizations such as PACs. Political influencers are present on Facebook and Google as well. TikTok also recently published its updated community guidelines \cite{tiktok_community}, but the guidelines do not mention how the platform moderates political content.



\subsection{Broadcast Regulations}

Federal law imposes disclosure requirements on political campaigns and broadcasters to ensure that the public can understand where campaigns spend money on reaching prospective voters and whether the broadcasters carry the ads in a non discriminatory manner. The rules for the broadcasters are set and administered by the Federal Communication Commission. In particular, the FCC's Political Programming staff oversees whether a broadcaster is favoring one candidate at the expense of the other by charging different rates or limiting the reach of candidate-sponsored ads \cite{federalcommunicationscommission_2022}. Specifically, the FCC's staff resolves issues related to the prohibition on censorship of candidate-sponsored ads; the “Lowest Unit Charges” and “Comparable Rates” that broadcasters charge candidates for their advertisements; and the on-air sponsorship identification for political advertisements. The FCC's staff also oversees the files that broadcasters must maintain for the public to easily access and inspect. 

In 2022, the FCC updated its regulations to require stations to maintain a files that contain the following information: (1) whether the broadcaster accepted or rejected the request to purchase broadcast time; (2) the rate charged for the broadcast time; (3) the date and time on which the communication is aired; (4) the class of time that is purchased; (5) the name of the candidate to which the communication refers and the office to which the candidate is seeking election, the election to which the communication refers, or the issue to which the communication refers; (6) in the case of a request made by, or on behalf of, a candidate, the name of the entity making the request; and (7) in the case of any other request, the name of the person purchasing the time, the name, address, and phone number of a contact person for such person, and a list of the chief executive officers or members of the executive committee or of the board of directors of such person \cite{fcc22}.

We extract an analogous evaluative criteria for online ads from these regulations for broadcasters that requires, at minimum, that the public should be able to evaluate how campaigns are spending money to target audiences, and whether the platforms are carrying the content in a non-discriminatory manner.

\subsection{Comprehensive Disclosures}

\begin{figure}[!h]
    \centering
    \includegraphics[width=0.47\textwidth]{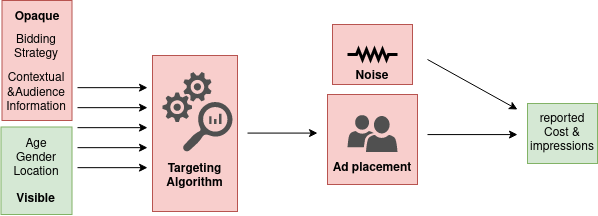}
    \caption{The visible and opaque aspects of online platform ad delivery mechanisms. Google discloses only the demographic segments targeted by the advertisers. Facebook reports only on the demographic segments that saw specific content. Hence, none of the platforms reveal full targeting and distributional parameters of ads.}
    \label{targeting_tools}
\end{figure}

Platforms have unique data about the political campaign's targeting parameters, how algorithms distribute or moderate content, and who actually saw the ads. But, as shown in figure \ref{targeting_tools}, platforms only make a fraction of that information available for public scrutiny. Typically, an advertiser runs an ad on the platform by selecting from a variety of targeting parameters, including age, gender, location, as well as some available specific contextual and audience properties. Google allows political advertisers to target based on demographic properties and specific contextual features (ad placements, topics, keywords against sites, apps, pages and videos) \cite{google_ads}. Facebook allows the use of demographic data for political ads, and also allows campaigns to use predefined lists of individuals or algorithmically generated ``look-alike'' audience lists \cite{facebook_political_ads}. After selecting targeting parameters, the advertiser chooses how it will pay for impressions, which the platform uses to calculate to whom the ad will be shown and at what cost. Given the advertiser’s choices, competing ads, and user properties, the platform uses complex algorithms to distribute the ad. It then creates reports for the advertiser about the number of impressions, as well as the total ad placement cost of the campaign. 

But the platforms’ transparency libraries do not contain the information they provide to advertisers. As discussed below, the appropriate disclosures for platforms should include information how their algorithms function. But even if we put information about their algorithms to one side, they should disclose to the public, at minimum, all the information they make available to advertisers about costs, impressions, and targeting parameters. Accordingly, we assess the effectiveness of the platforms' ad libraries by comparing what is disclosed currently against what could be made available to a hypothetical advertiser on that platform.


For a platform's ad moderation practices there is little precedent to draw on to develop standards for appropriate disclosures. We examine whether the public can understand if the policy has been applied consistently and if the platform has provided an adequate explanation for its decision to moderate an advertisement.

\section{Data} 

We create a large scale dataset of political ads and content for Facebook, Google (including YouTube), and TikTok. The dataset contains more than 850,000 ads on Facebook, Google, and YouTube and 2,7 million political videos on TikTok. We focus on content that was created up to two months prior to the US Elections. For ads, we collect who sponsored them, what specific targeting parameters \& audience characteristics were used to distribute them, as well as their reach in terms of impressions and the corresponding cost. We also crawl the Facebook \& Google ad libraries to locate moderated ads. For TikTok, we generate a list of political hashtags, which we use to crawl videos from the platform. We also use a list of political influencers \& Hypehouses (40 democrats, 56 republicans), and download all their videos. For each video, we also collect the corresponding engagement metrics in terms of views, likes, shares, and its description. We also collect video creators' metadata. Furthermore, we locate which TikTok videos were assigned an election-related warning by the platform, since the platform soft-moderated election related content, by placing a warning banner saying \textit{Get info on the U.S. elections}, The banner was linking to a guide including authoritative information about the election process. An in-detail description about the collected dataset can be found in table \ref{data_overview} and in appendix \ref{data_appendix}.

\section{Methods}
First, we quantify the prevalence of the political ads on the platforms. We quantify the number of impressions for political ads on Facebook, Google, and YouTube, and the cost to place this content. For TikTok, we document the number of views for videos produced by political influencers and of videos containing political hashtags gather in our dataset (Figure \ref{overview}).

\begin{table*}
   \caption{Overview of the collected dataset. We provide information about how we obtained the data, for which periods, what attributes, and what post-processing we performed. A more in detail explanation can be found in appendix \ref{data_appendix}.}
\label{data_overview}
\resizebox{1\textwidth}{!}{
\begin{tabular}{l|l|l|l}
                                                                                    & \textbf{Facebook}                                                                                                                                                                                                                                         & \textbf{Google}                                                                                                                                                                                                            & \textbf{Tiktok}                                                                                                                                                                                                                           \\
\textbf{Source}                                                                     & FB ad library                                                                                                                                                                                                                                             & Google ad archive                                                                                                                                                                                                          & Website                                                                                                                                                                                                                                   \\ \hline
\textbf{Collection method}                                                               & Ad library API                                                                                                                                                                                                                                            & Crawled the ad archive                                                                                                                                                                             & Crawled the website                                                                                                                                                                                                                       \\ \hline
\textbf{Scope}                                                                      & \begin{tabular}[c]{@{}l@{}}Ads related to politics by advertisers who spent\\  at least \$100k\end{tabular}                                                                                                                                               & \begin{tabular}[c]{@{}l@{}}Political ads placed through Google ad services\\   by advertisers who spent at least \$100k\\ (includes ads appearing in Google search,\\  third party websites, and on YouTube.)\end{tabular} & \begin{tabular}[c]{@{}l@{}}(All) videos by creators who had trending\\  content related to political hashtags\\ Videos created by popular influencers and\\  hypehouses engaged in political\\  campaigning (40 dem; 56 rep)\end{tabular} \\ \hline
\textbf{Date range}                                                                 & \begin{tabular}[c]{@{}l@{}}Sep 1 - Nov 4 (60 days preceding the election)\\ No ads in the week preceding the\\  election (FB enforced ban)\end{tabular}                                                                                                   & Sep 1 - Nov 4 (60 days preceding the election)                                                                                                                                                                             & \begin{tabular}[c]{@{}l@{}}Sep 1 - Oct 15 (limited by technical\\  restrictions; see Appendix B)\end{tabular}                                                                                                                             \\ \hline
\textbf{Attributes}                                                                 & \begin{tabular}[c]{@{}l@{}}For each ad:\\ Sponsored or not\\ Who sponsored it\\ Content\\ Audience data\\ Upper/lower bounds of generated impressions\\ Upper/lower bounds of generated cost\\ Age, gender, location stats of who saw the ad\end{tabular} & \begin{tabular}[c]{@{}l@{}}For each ad:\\ Content\\ Upper/lower bounds of generated impressions\\ Upper/lower bounds of generated cost\\ Age, gender, location targeting criteria\end{tabular}                             & \begin{tabular}[c]{@{}l@{}}For each creator:\\ Account description\\ \# followers\\ General popularity\\ For each video:\\ \# views, likes, shares\\ description\end{tabular}                                                             \\ \hline
\textbf{Dataset size}                                                               & \begin{tabular}[c]{@{}l@{}}749,556 ads\\ 803 advertisers\\ (65\% of political ads in the specified period)\end{tabular}                                                                                                                                   & \begin{tabular}[c]{@{}l@{}}117,607 ads\\ 490 advertisers\\ (83\% of political ads in the specified period)\end{tabular}                                                                                                    & \begin{tabular}[c]{@{}l@{}}2,690,923 videos\\ Over 61,000 creators\end{tabular}                                                                                                                                                           \\ \hline
\textbf{\begin{tabular}[c]{@{}l@{}}Moderated\\ ads \& content: method\end{tabular}} & \begin{tabular}[c]{@{}l@{}}Crawled ads in dataset to look for FB’s\\  removal flag\end{tabular}                                                                                                                                                           & \begin{tabular}[c]{@{}l@{}}Crawled the archive again in January and compared to\\  initial crawl to locate missing ads\end{tabular}                                                                                        & \begin{tabular}[c]{@{}l@{}}Looked for Tiktok videos having\\  an election-related’ warning\end{tabular}                                                                                                                                   \\ \hline
\textbf{\begin{tabular}[c]{@{}l@{}}Moderated\\ ads \& content: size\end{tabular}}   & \begin{tabular}[c]{@{}l@{}}8,635 ads\\ 253 advertisers\end{tabular}                                                                                                                                                                                       & \begin{tabular}[c]{@{}l@{}}9,735 ads (content available through initial crawl)\\ 451 additional ads (not in initial crawl;\\  content unavailable)\end{tabular}                                                            & \begin{tabular}[c]{@{}l@{}}243,440 TikToks having an\\  election related warning\end{tabular}                                                                                                                                             \\ \hline
\textbf{\begin{tabular}[c]{@{}l@{}}Post-\\ processing\end{tabular}}                 & \multicolumn{2}{l|}{\begin{tabular}[c]{@{}l@{}}Extract text using Google Cloud Vision and speech-to-text APIs\\ Match advertisers to registered political entities in FEC database\\ Match advertisers to records in FollowTheMoney to categorize organization\\  type and type of content created\end{tabular}}                                                                                                                                                                        & \begin{tabular}[c]{@{}l@{}}Manually categorized HypeHouses based \\ on organization type and content  type \end{tabular}   \\ \hline                                                                                                                 
\end{tabular}}
\end{table*}

Second, we analyze whether the data provided by the online platforms adequately explain the platforms' decisions and algorithms using the three analytical criteria described  in section \ref{evaluative}. Table \ref{overview_methods} provides an overview of the methodologies we describe in detail next, together with the corresponding evaluative criteria and the sections we report our related results.

\begin{table*}
\caption{Overview of the different methods we use for evaluating platforms' given the evaluative criteria we developed. We use a uniform heading scheme across methods \& results to efficiently connect the two sections. }
\label{overview_methods}
\resizebox{1\textwidth}{!}{\begin{tabular}{llll}
\textbf{ Method  }                                                                                                                                                                                                                                                                         & \textbf{ Platform    }                                                  & \textbf{ Evaluative Criterion }                                                                                                                                                                                                                                                                   & \textbf{ Section in methods \& results }                                                                                                                                \\ \hline
\begin{tabular}[c]{@{}l@{}}\textbf{1.} Quantification and qualitative evaluation of the \\ information  provided in the ad libraries by\\  assessing Biden’s and Trump’s ad campaigns\end{tabular}                                                                                           & \begin{tabular}[c]{@{}l@{}}Facebook \\ \& Google\end{tabular} & \begin{tabular}[c]{@{}l@{}} \textbullet\ Do the disclosures meet \\ self described objective (A)?\end{tabular}                                                                                                                                                                                & Assessing information in the ad libraries                                                                                                                     \\ \hline
\begin{tabular}[c]{@{}l@{}}\textbf{2.} Reverse engineering the platforms’ targeting\\ algorithms by placing dummy ads \\ on the advertising networks\end{tabular}                                                                                                                            & \begin{tabular}[c]{@{}l@{}}Facebook \\ \& Google\end{tabular} & \begin{tabular}[c]{@{}l@{}}\textbullet\  Do the disclosures meet \\ self described objective (A)?\end{tabular}                                                                                                                                                                                & \begin{tabular}[c]{@{}l@{}}Reverse engineering the platforms’\\ targeting algorithms.\end{tabular}                                                            \\ \hline
\begin{tabular}[c]{@{}l@{}} \textbf{3.} Regression modeling connecting targeting \&\\  audience characteristics to ads' cost per\\  impression\end{tabular}                                                                                                                                   & \begin{tabular}[c]{@{}l@{}}Facebook \\ \& Google\end{tabular} & \begin{tabular}[c]{@{}l@{}}\textbullet\ How do disclosures \\ compare to broadcast (B)?\end{tabular}                                                                                                                & \begin{tabular}[c]{@{}l@{}}Connecting targeting \& audience \\ characteristics to ads’ cost per impression.\end{tabular}                                      \\ \hline
\begin{tabular}[c]{@{}l@{}} \textbf{4.} Identification of ad instances that their targeting \\options on Google match with ad instances’ \\distributional information on Facebook, assesing how \\algorithms  deliver ads based on specific targeting \\choices  of the advertisers.\end{tabular} & \begin{tabular}[c]{@{}l@{}}Facebook \\ \& Google\end{tabular} & \begin{tabular}[c]{@{}l@{}}\textbullet\ Do the platforms disclose\\ all that they know (C)? \end{tabular}                                                                 & \begin{tabular}[c]{@{}l@{}}Crossplatform comparison of targeting\\  parameters \& audience characteristics.\end{tabular}                                      \\ \hline
\begin{tabular}[c]{@{}l@{}} \textbf{6.} Locating and quantifying the prevalence and\\  reach of specific ads that were partially\\ moderated by the platforms\end{tabular}                                                                                                                    & \begin{tabular}[c]{@{}l@{}}Facebook \\ \& Google\end{tabular} & \begin{tabular}[c]{@{}l@{}}\textbullet\  Do the disclosures meet \\ self described objective (A)? \\ \textbullet\ Do the platforms disclose all\\ that they know (C)?\end{tabular} & \begin{tabular}[c]{@{}l@{}}Characterizing the magnitude and\\  effectiveness of moderation for Facebook\\  \& Google\end{tabular} \\ \hline

\begin{tabular}[c]{@{}l@{}} \textbf{7.} Regression modeling connecting TikTok video \\ properties and its moderation\end{tabular}                                                                                                                                                             & TikTok                                                        & \begin{tabular}[c]{@{}l@{}}\textbullet\ Do the platforms disclose all\\ that they know (C)?\end{tabular}                                                                 & \begin{tabular}[c]{@{}l@{}}Connecting TikTok video properties to \\ their moderation.\end{tabular}                                                            \\ \hline
\end{tabular}}
\end{table*}



\subsection{Distribution of political content}

\subsubsection{Assessing information in the ad libraries}For the platforms that have political ad libraries (Facebook, Google, \& YouTube), we assess the platforms' role in shaping the distribution strategy. Since platforms do not provide all targeting information associated with an ad, we explore what the limited data can tell us about how the platforms' tools were used to target specific audiences. Specifically, we quantify the unique number of ads Biden's and Trump's campaigns placed in terms of content, location, age and gender demographics. We also locate how the same ads were distributed across different platforms, and we compare the distribution metrics provided by Facebook and Google to assess what information ad libraries can provide.

\subsubsection{Reverse engineering the platforms’ targeting algorithms.}Next, we attempt to reverse engineer the platforms' targeting algorithms. We do that by creating advertising accounts on Facebook and Google, and evaluating the cost and impression estimates for hypothetical ads that mimic the targeting criteria of original political ads that ran on the platforms. If the cost/impression estimates for the hypothetical ads deviate significantly from the reported ranges for the ads that did run, we assume that advertisers used additional targeting options. 

Specifically, we create four different dummy ads to investigate the relationships in more detail. On Google, our dummy ad targets the whole of the United States and to all available genders and ages, and we calculate the upper and lower impressions it will generate for a budget varying from \$10, to \$1,000,000. We do the same for an ad on YouTube targeting Pennsylvania and females between 25-34. For Facebook we target one ad at the whole of the United States and all genders and ages. Lastly, our other dummy ad on Facebook targets California, and females of all ages. Based on the reported upper and lower values, we interpolate cost and impressions and calculate an area of plausibility. If an ad in the ad libraries with the same targeting options falls within this area, it suggests that the advertisers actually used these targeting options. In case that an ad falls outside the area, it suggests that ads might have used additional targeting options that platforms did not disclose. In this way we can assess information quality in the ad libraries. One notable limitation of this approach is that we ran these dummy ads at a different time period from the political ads we examined. As a result, the analysis is illustrative of the technique and should not be taken as definitive. 

\subsubsection{Connecting targeting \& audience characteristics to ads' cost per impression.}To further explore how algorithms distribute political content, we investigate the sensitivity of cost/impressions ratios to different audience characteristics and targeting properties. We create models that analyze ads generated by the Biden and Trump campaigns, and uncover how ad specific properties link to ads' distribution. For Facebook, we create a linear regression model that has as dependent variable the cost/impressions ratio, and as independent the ratio of individuals for each state that viewed an ad, the ratio of individuals that were either male or female and belonged to the age buckets 18-24, 25-34, 35-44, 45-54, 55-65,65+, and whether the ad was placed by the Biden or Trump campaign. Since the Google Ad Archive aggregates cost and impression into intervals, we create an ordinal logistic regression model that has dependent variable the cost of an ad, and as independent the generated impressions, whether the ad was text (ad on google search), image (ad on third party affiliates), or video (YouTube ad), the targeted genders (male, female), different age groups (18-24, 25-34, 35-44, 45-54, 55-65, 65+), the magnitude of region targeting (USA, state level, county level, zipcode level), and whether the ad was placed by the Biden or Trump campaign. Based on located associations, we uncover factors that shape the algorithmic distribution of content. 

\subsubsection{Crossplatform comparison of targeting parameters \& audience characteristics.}To understand what full disclosures can tell about the algorithms that distribute political advertisements, we use as data the cross-platform tactics of advertisers. Following the principles of personalized advertising, we make the strong assumption that advertisers would have targeted specific demographics with the same content across platforms, in order to maximize their influence potential. Therefore, we identify 35 unique image ad designs created by Biden \& Trump that correspond to 12,448 unique ad placements on Facebook and 3,055 ad placements on Google. Similarly, we identify 72 unique video ad designs that correspond to 13,840 ad placements on Facebook and 4,383 ad placements on Google. We then identify ad instances that their targeting options on Google match with ad instances' distributional information on Facebook, and assess how algorithms deliver ads based on specific targeting choices of the advertisers. We do so by focusing on two sets of ads, namely YouTube ads placed by the Biden campaign to all genders and image ads placed by the Trump campaign to all available age groups. In this way, we uncover how consistent is Facebook's algorithmic distribution.

\subsection{Moderation of Political advertising and
content} 

We analyze platform tactics in algorithmic ad \& content moderation based on each platform's policies.

We use two methods to evaluate TikTok's moderation of election related content. 

\subsubsection{Connecting TikTok video properties to their moderation}
Our second technique investigates TikTok's use of warning labels related to the U.S. elections that uses a logistic regression model to predict whether a video was flagged. We use as independent variables the likes, shares, comments, and views it generated, the average amount of likes the video's author collected, as well as the presence of three election-related (\#biden,\#trump, \#vote) and three non-election related (\#blm, \#abortion, \#gun) hashtags. Furthermore, we calculate the ratio of political videos flagged for each user in our dataset. Based on the results of both analyses, we can uncover features that constituted algorithmic content moderation. 

\subsubsection{Characterizing the magnitude and effectiveness of moderation for Facebook \& Google}
For Facebook, Google, and YouTube, our investigation of moderation practices examines the ad libraries to document how many ads were moderated and how many individuals saw ads that were flagged. In addition, we manually reviewed a set of 200 moderated ads for each platform, which allows us to qualitatively understand features of the moderation process.  Furthermore, based on the set of moderated ads in our sample, we locate other ads that contained the same content, but were not moderated. We quantify their distribution on the platforms, and assess the robustness and degree of explainability of the moderation process. These three steps allow us to uncover patterns about who was moderated, why, and how effective was this moderation. 

\section{Results}

\subsection{Political content distribution} \label{distribution}

\subsubsection{Assessing information in the ad libraries: Platforms provide limited insights about how campaigns used their targeting tools.}

Table \ref{ad_targeting} presents the demographic distribution of Biden and Trump ads on Google and Facebook. This superficial view seems to suggest that advertisers rarely resorted to micro-targeting, and instead applied broad criteria to target general segments of the society. For Google, which publishes targeting data, we see that both Biden and Trump appear to infrequently use the platform's fine-grained demographic targeting (except for targeting by state). Facebook, provides audience characteristics of individuals who saw the ads, rather than targeting choices, but the pattern is similar.


\begin{table}
\caption{Demographic distribution of content by candidate on Facebook \& Google. For Google we show the targeting choices of advertisers. For Facebook we report who was shown an ad (audience characteristics).}
\resizebox{0.45\textwidth}{!}{\begin{tabular}{r|cc|cc}
&  \multicolumn{2}{c}{Google} & \multicolumn{2}{|c}{Facebook} \\ \hline
Distribution Strategy                 & Biden                & Trump        & Biden                & Trump            \\ \hline
Age                   & 3\%                  & 0\%           & 34\%                  & 31\%                        \\
Gender                & 1\%                  & 0\%           & 6\%                  & 7\%                        \\
zip Code/County              & 17\%                 & 19\%       & -                  & -                           \\

State              & 81\%                 & 69\%       & 87\%                  & 96\%                           \\
Age \& Gender \&  zip Code/County  & 0\%                 & 0\%      & -                 & -                           \\
Age \& Gender \&  State  & 1\%                 & 0\%      & 6\%                  & 7\%                            \\
\end{tabular}}
\label{ad_targeting}
\vspace{-8mm}

\end{table}

Yet, a closer look of the actual content distributed on the platforms contradicts this superficial view.
 For example, we used algorithmic tools \cite{cld2} and manual coding, to detect all ads placed by Biden and Trump in the Spanish language. On Google, the majority of the 1724 ads we located that were in Spanish language had no demographic targeting and were sent to geographies that did not have large Hispanic populations. Even for zip code targeted ads, the percentage did not exceed on average 30\% on Google and 12\% on YouTube. In other words, the campaigns potentially used some undisclosed contextual targeting criteria to place the ads. On Facebook, we also located 626 ads in Spanish, but since the ad library provides distributional data only at the state level, we were not able to evaluate how they were targeted.


\subsubsection{Reverse engineering the platforms’
targeting algorithms.} 
\begin{figure*}[!htb]
    \centering
    \includegraphics[width=1\textwidth]{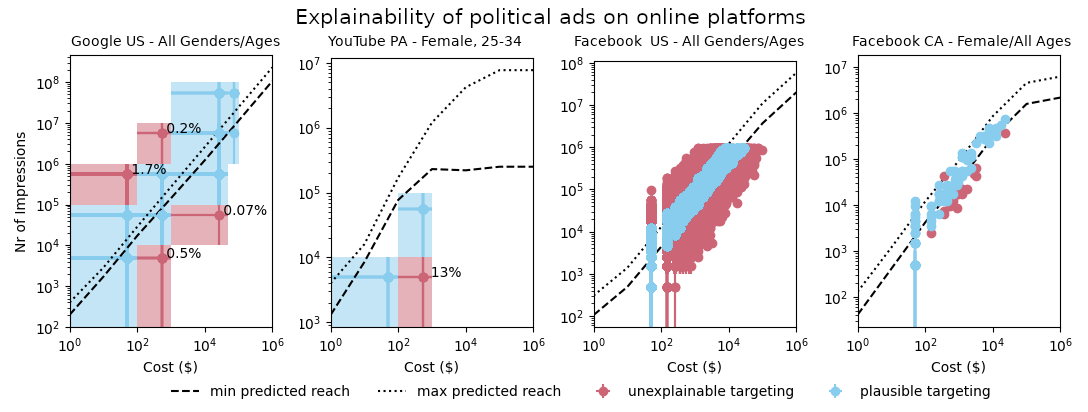}
    \vspace{-7mm}
    \caption{We plot the generated cost per impression of ads in the ad-libraries that were (1) targeted to all genders \& ages on Google, (2) to Females, between 25-34 on YouTube, (3) were seen by all genders \& ages in the US on Facebook, and (4) only by females of all ages located in California on Facebook.  For Facebook, lower \& upper bounds are provided for the impressions. For Google, lower \& upper bounds are provided for cost \& impressions, given the extensive “bucketing” of the parameters performed by the ad libraries when reporting them, which are denoted in the figures with boxes. Points represent the median value of the boxes. We compare the generated cost-per impression of ads with the cost-per impression of a set of dummy ads we placed on the platforms with the exact same targeting parameters \& audience characteristics. Black lines represent the upper and lower boundaries of an ad’s cost-per-impression as we extracted them from the dummy ads. We label an ad placement as ``plausible targeting'', when the ad cost-per-impression overlaps with the one we calculated, denoting that we can assume that the ad library provides all relevant targeting parameters/audience characteristics about an ad.  Similarly, an placement labeled as ``unexplainable targeting''  represents an ad whose cost-per-impression is outside the upper and lower reach values that we calculated, meaning that potentially platforms do not disclose full information about the distribution of the ad.
}
    \label{obscurity}
\end{figure*}

Similarly, our reverse engineering of the platforms' tools evidence indicates that the campaigns used undisclosed targeting strategies. Figure \ref{obscurity} presents how the distribution of the actual ads in our sample falls within the retrieved cost-per-impression boundaries. We find that the reverse engineering data do not always correspond to the targeting/distributional information provided in the ad libraries. For Google, this discrepancy is small, when looking at ads distributed over the United States across all genders and ages, with only about 2\% of the ads reported in the ad libraries following out our calculated boundaries. This discrepancy is significantly larger when comparing YouTube ads placed in Pennsylvania to Females between 25 and 34, with the disagreement reaching 13\%. However, the true discrepancy may be much larger because Google uses very large reporting buckets for costs and impressions, as Figure \ref{obscurity} reveals. By way of illustration, an ad is assigned the same value of $\leq$ 10,000 impressions and $\leq$\$100 costs whether it is shown to 500 individuals at a cost of \$5 or 9,000 individuals for \$90. 
\setlength{\belowcaptionskip}{-10pt}
\setlength{\belowcaptionskip}{0pt}

We locate similar discrepancies when reverse-engineering the information for Facebook. For ads placed to Females of all ages in California, the disagreement between cost/impression data from our analysis and from the ad libraries is 14\%. For ads placed in the United States generally, across all genders and ages, the disagreement exceeds 27\%. These results illustrate that the available information ad libraries provide is not sufficient to understand how political advertisers used the platform to distribute political messages. Even if the detected discrepancies are a result of specific undisclosed parameters that influence targeting (e.g. auction system, time of placed ad, etc.), the disclosures do not meet the platforms' self-described objective of making political advertisers accountable. Although the ad libraries provide information about the content of ads, they do not allow the public to understand the exact segments of the society advertisers wanted to target, the price they paid, or  how  algorithms transformed these intentions to a specific content distribution.

\subsubsection{Connecting targeting \& audience 
characteristics to ads’ cost per impression: Disclosures and targeting algorithms fall short compared to existing broadcasting policies.} \label{distribution2}
Our regression analysis results (tables \ref{regression1} \& \ref{regression2}, appendix) reveal specific shortcomings for algorithms and platform disclosures, both for Facebook and Google, since we discover that there was no parity in the cost of ads between advertisers. We also see that different demographic targeting and audience characteristics resulted in different ad placing cost. 

For Facebook, there was no difference in the number of impressions per dollar an ad generated, regardless of the gender of individuals who saw them. Nevertheless, age was a factor associated with different amounts of generated impressions. Specifically, placing ads to older populations (> 55) and very young populations (18-24) was significantly more expensive than placing ads to individuals between 25 and 54. Furthermore, ads cost varied between different states. For example, the most expensive impressions were found in the states of Massachusetts, Rhodes Island, and Washington DC, while the cheapest impressions were generated in the states of Idaho, Kentucky, and Mississippi. Interestingly, ads that Trump placed were overall more expensive by impression compared to those of Biden. 

On Google and YouTube, 
as with Facebook, there was a difference when targeting different age demographics. Targeting individuals between the ages of 18-24 was by far the cheapest, while the most expensive targeting groups were people aged between 25-34. The more location-specific the targeting, the more expensive was the ad placement, with zip code targeted ads being the most expensive and US-general ads being the cheapest. In contrast to Facebook, the ads that Trump placed were cheaper than those placed by Biden. 

Comparing these results to the broadcast regulations we observe multiples instances of a lack of rate parity between campaigns and across different strategies. Of course, these rate differences can be attributed to multiple factors, such as additional contextual targeting criteria, which where not provided by Facebook or Google, as well as further information about how their algorithms distribute and price content. But it is worth noting that the potential for a broadcaster to favor one campaign over another led to the FCC rules on parity between campaigns and commercial advertisers, to ensure that the broadcaster was acting appropriately. Another issue is that the algorithmic promotion and demotion of content by the platforms runs into the concern that the intermediary might covertly limit the reach of candidate-sponsored ads. But our analysis shows that there are unexplained artifacts in the distribution that can be tied back to the algorithmic choices of the platforms.


\subsubsection{Cross-platform comparison of targeting parameters \& audience characteristics: Full disclosures reveal properties of algorithmic distribution} \label{distribution3}

 Building on our results from the prior sections, we evaluate what the cross-platform tactics of advertisers can reveal about how content was algorithmically distributed to the public. We do this by pairing ads that were shown on Google and Facebook so that we obtain data about targeting criteria from Google and the actual distribution from Facebook. 

\setlength{\belowcaptionskip}{-10pt}
\begin{figure}[!h]
    \centering
    \includegraphics[width=0.47\textwidth]{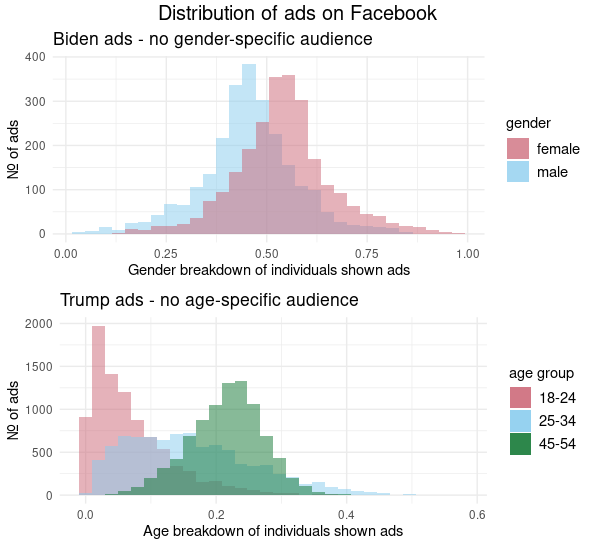}
        \vspace{-3mm}
    \caption{Distribution of ads on Facebook, that were matched with their targeting parameters on Google. We show how Biden ads targeted to both genders were distributed on Facebook (up). We also show how Trump's image ads that were targeted to the whole age spectrum were distributed among three age groups. For Gender, we find that ads distribution had statistical parity given the user demographics on the platform. For age, we find that ads were not distributed in rates that correlate with the Facebook's age demographics in the US.}
    \label{facebook_distribution}
\end{figure}
\setlength{\belowcaptionskip}{0pt}

\setlength{\belowcaptionskip}{-10pt}
Figure \ref{facebook_distribution} shows how two specific sets of ads on Facebook were distributed across genders and age-groups respectively, given that they were not targeted to specific gender or age subgroups on Google. For video ads placed by Biden, which were not gender targeted, we find that
they were distributed unequally among males and females. On average, ads were shown 45\% to males, and 53\% to females, while the standard deviation of the distribution for each gender was 12\%. This discrepancy about the ads can be potentially attributed to the gender demographics that used Facebook in the US in 2020, which was 54.5\% for females and 43.5\% for males \cite{napoleoncat}.

By analyzing image ads for Trump, that did not have any limitations in age targeting, we uncover similar patterns. Focusing on three age-groups, we find that on average an ad would be shown at a rate of 6\% to individuals between 18-24, at a rate of 16\% to individuals between 25-34, and at a rate of 21\% to individuals between 45-54. In contrary to gender distribution, these values do not correspond to the Facebook user demographics in 2020, which were 8\% for individuals between 18-24, 13\% for individuals between 25-34, and 7\% for individuals between 45-54. This means that algorithmic targeting resulted in a disparate distribution of ads across age groups. These findings provide additional support for the public interest in further understanding how the distribution of the campaigns ads was affected by the platforms' algorithmic choices.

\subsection{Moderation of political advertising and content} \label{moderation}

Focusing on moderation, we evaluate how platforms disclose their choices and practices when removing political ads (Facebook, Google, YouTube), and handling political content (TikTok).

\subsubsection{Characterizing the magnitude and
effectiveness of moderation for Facebook \& Google: Unexplainable ad moderation practices.}

Our study documents how difficult it is to understand how the platforms apply their moderation policies to political ads.

On Google and Facebook we see that a large number of ads across a wide range of advertisers were removed (table \ref{removal_ads}, appendix). Google removed 13.3\% of the political ads from its network, YouTube removed 4.5\%, while Facebook only 1.2\%. Despite their removal, these ads generated a significant amount of user impressions. Furthermore, these decisions affected a significant number of advertisers. Google removed at least one ad from 256 advertisers (18\% of all), YouTube at least one ad from 307 advertisers (22\% of all), and Facebook from 266 advertisers (31\% of all). 

\begin{figure*}
    \centering
    \includegraphics[width=0.96\textwidth]{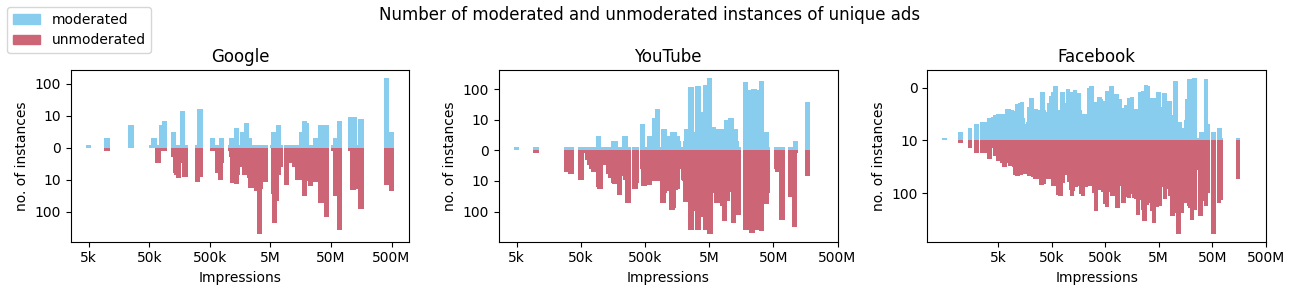}
            \vspace{-3mm}
    \caption{  Comparison of different instances of moderated ads across platforms. The light blue bars show how many instances of a single ad were moderated, and maroon bars show how many instances of the same ad were not. Results suggests an inconsistent moderation of content across platforms, with some instances of the same ad being removed and some others not.}
    \label{crossplatform_moderated}
\end{figure*}

Figure \ref{crossplatform_moderated} illustrates how different instances of the same ad design were moderated. For each platform we find a significant number of ads that contained some instances where the ad was removed, and some that were not. On Facebook, this appeared in 51\% of the moderated ads. For Google, the amount of these ads was 75\%, while for YouTube it was 65\%. In total, we found 11,549 ad instances across platforms that were not moderated, although at least one identical to them was removed. In median for Google, non-moderated ad instances resulted in the generation of 1.1 billion impressions, compared to 700 million for the moderated ones. For YouTube, non-moderated ad instances generated 1.2 billion impressions, while moderated instances 900 million. For Facebook, these numbers were 440 million and 200 million respectively. These results suggest that inconsistent ad moderation had serious implications, since both moderated ads resulted in a significant amount of impressions, and also their unmoderated counterparts resulted in an even higher diffusion of problematic content. Furthermore, the platform ad libraries do not provide any explanation why and when an ad was removed, therefore, is not possible to assess the reasons for these discrepancies. Especially for Facebook, we find that even the classification of ads as removed was inconsistent. When manually reviewing a random sample of 200 moderated ads, we found that 35 were no longer labeled as removed. These results raise questions about the efficacy, robustness, and explainability of the moderation practices.


\subsubsection{Connecting TikTok video properties to their moderation} Focusing on content moderation on TikTok, we find 505,062 videos that contained at least a hashtag from our curated hashtag list. From these, 243,440 videos (48.2\%) were labeled with a U.S. election warning. We find that election related content was a driver for this moderation. Figure \ref{tiktok_moderated}-Left (Appendix) shows the logistic regression results for predicting whether a TikTok video containing political hashtags has been labeled or not. We find that election related hashtags (Biden/Trump/Vote) were strong predictors for content moderation, while non-election related hashtags were not as much (e.g. BLM, abortion, gun). Similarly, video views and likes seem not to have been associated with the probability of a video being labeled, while both video creator popularity and sharing/commenting were. The more popular the content creator, the less likely that a video of theirs was going to be labeled. Furthermore, the more organic interactions a video gathered in terms of shares and comments, the more likely it was to be labeled. Analyzing the distribution of moderated videos by author (figure \ref{tiktok_moderated},right), we find a discrepancy in the videos labeled by user, with some videos containing election-related hashtags of a user being flagged, and some not. These results illustrate that although the content of videos was strongly associated with warning placement, there is a need for additional information about TikTok's practices. For example, it would be important to know the exact definition of ``election-related'' on the platform, or whether features such as video description or author information were taken into consideration for labeling content.
\subsection{Consequences of different definitions of "political ads"}
\setlength{\belowcaptionskip}{0pt}

In evaluating what self-disclosures reveal about advertisers and algorithms, we found that each platform's defined political ads differently, which made it difficult to make comparisons across platforms. As shown in Figure \ref{FEC} (Appendix), each platform has a different definition of what is a political ad, which dictates whether and to what extent an advertiser will be allowed to use platforms' targeting tools and whether their campaigning information will be included in the ad libraries. Google bases its definition solely on the identity of the advertiser, while Facebook takes a more expansive view that also captures advertisers running issue-based ads. As a result, we see that the Facebook ad library has a higher proportion of advertisers who are not registered with the FEC as compared to the Google Ad Archive (45\% vs 25\%). In contrast, the narrower definition of political ads on Google does not allow us to see the targeting activities of a broad set of advertisers, including a significant amount of NGOs (23\% on Facebook vs 8\% on Google).  In \textit{Citizens United v. FEC}, the United States Supreme Court upheld the disclosure requirements for issue-based ads on broadcast stations because "[e]ven if the ads only pertain to a commercial transaction, the public has an interest in knowing who is speaking about a candidate shortly before an election." That same public interest here should require disclosures about all ads involving ``electioneering communications'' \cite{fec_2022}. Platforms should not be allowed to strategically limit what information becomes available to the public.

In the case of TikTok, we see that while direct political advertising is prohibited, there are other forms of political campaigning on the platform. We looked more closely at the 96 influencers we identified to uncover links to political entities, especially those who had registered with the FEC. In our sample, we found six influencers directly linked to political NGOs, two to PACs, four to political merchandise, while 18 were asking for donations. Given the importance of this type of influencer advertising on of social media platforms, future work should investigate the funding mechanisms for influencer-driven political advertising in more detail \cite{mathur2020manipulative}.

\section{Policy recommendations}
As our work demonstrates, there is an urgent need to standardize disclosures across platforms. In a white paper, Edelson et al. \cite{edelson2021universal}, outline a useful technical standard for universal digital transparency. They propose the creation of a public repository and a technical standard that will make advertisers and platforms accountable for disclosing how they distribute content to the public. Our study builds on that work by demonstrating how the current voluntary disclosures about political ads by the platforms fall short of providing meaningful insights, and why such a standard is necessary for political advertising.

Specifically, we recommend that the FEC (or alternative federal agency) creates a cross-platform database that provides information about advertisers and platforms. Such a database should be accompanied by a standard that provides a universal identifier for each advertiser, so their campaign activities can be tracked across different platforms. In our study, the mapping of entities was a time consuming and complicated process. Furthermore, the standard should clearly define what is considered "political," so that there is an equal baseline for disclosures across the platforms. Such definition should be carefully chosen, given the complexity and non-triviality of the concept \cite{settle2018frenemies}.  The same applies for available targeted information. Platforms should disclose both the full and detailed targeting and distribution parameters (audience characteristics) of ads, since anything less than this results in an incomplete and inefficient evaluation of advertisers' campaigns and platforms' decisions. Finally, the full disclosures of targeting criteria can facilitate understanding the specific campaigning techniques attempted to influence voters. 

We believe that the creation of a standard and repository should be accompanied with detailed regulations that protect the public and ensures fairness among political advertisers. Drawing from the broadcasting regulations, we observed an apparent difference in rates between different advertisers. Platforms should disclose how they algorithmically control the price and reach of content, whether platforms deliberately or unintentionally limit the reach of candidate-sponsored ads, how they ensure parity, and provide information that can reveal whether advertisers or platforms target segments of the society in a biased way. Similarly, since we discovered significant inconsistencies in ad moderation, we argue that platforms should be obligated to disclose when and why an ad was removed, and make the removed content available for review in a neutral repository. This can make platforms accountable for their decisions and algorithms, and can ensure the fair moderation of content among advertisers. 

Lastly, the broad reach of the influencers that we document on TikTok highlights the need for regulations that require disclosures about their sources of funding and other activities. And such influencer-driven marketing is also present on Google and Facebook properties, but is not disclosed in their ad libraries. Because Federal Trade Commission's endorsement guidelines are designed for commercial transactions and not political campaigning, this is an area where the Federal Election Commission would need to develop comprehensive disclosure requirements.
\section{Conclusion}

In this study, we evaluated what the platform disclosures could tell the public about their role in the distribution and moderation of political advertising. By taking the political ad libraries and platforms transparency mechanisms seriously, we undertook large scale data analysis of political ads on Facebook, Google, and TikTok. Our study demonstrated the existence of strong barriers to public understanding advertisers' tactics. We also found evidence that platforms' disclosures falls well short of what is required under the law for broadcasters. Finally, we showed why we meed more accurate and comprehensive disclosures to understand and robustly evaluating targeting tools and algorithmic moderation.     

\section{Acknowledgments}

This study was supported by a Princeton Data Driven Social Science Initiative Grant. We thank Laura Edelson, Andy Guess, and Matt Salganik for constructive feedback on the final manuscript. We are also grateful for the early feedback from the research seminar run by Princeton's Center for the Study of Democratic Politics and later feedback from the MPSA'22 panel on political marketing. We would also like to thank Eli Lucherini for support in data analysis, Ashley Gorham for conceptual contributions in the early stages of the project, Juan Carlos Medina for a part of the data collection, and Milica Maricic for support in classifying political ads and creating the website of the project.

\bibliography{acmart}

\appendix
\onecolumn
\section{Ethical Considerations} 

\subsection{Data selection}
Facebook provides, through special agreements for researchers, access to the FORT dataset \cite{facebook_fort}, which purports to contain more detailed information about ad targeting on the platform as compared to the public library we analyzed. We decided not to use the FORT data for two reasons. First, we wanted to focus on data that was available to the public at large. Second, at the time we conducted our analysis, the platform did not provide us with appropriate assurances that we 
could use that data without any research \& publication restrictions. 
\subsection{Privacy concerns}
The analysis of the Facebook and Google data included only public information about ads and advertisers, as collected from the their APIs. For TikTok, we crawled the platform and collected only the public meta-data of user-generated TikToks, and not the actual videos. For security purposes, only one researcher of the project has access to the information, and will delete it upon completion of the study, as proposed by ethical data collection \& analysis guidelines \cite{franzke2020internet}.

\section{Data} \label{data_appendix}

\subsection{Facebook}
Facebook's Ads Library contains information about ads related to politics, credit, housing, and employment. The platform provides information about whether an ad was sponsored or not, who sponsored it, its content, and data about the audience. Specifically, it provides a lower and upper number of generated impressions and cost, as well as which user groups \textbf{saw} the ad in terms of age, gender\footnote{Gender is a spectrum. Nevertheless, both Facebook \& Google use a binary classification of genders. We adopt this language for the specific analysis, but we disagree with this form of classification.}, and location. Using the Ads Library API service, we collected all political ads placed in the 60 days leading up to the election (September 1st to November 4th), which tracks the legal definition of "electioneering communications," by advertisers who spent at least hundred thousand dollars. Our final dataset consisted of 749,556 ads created by 803 advertisers, and we collected it in November 2020. This represented approximately 65 \% of political ads in the specified period. Facebook also enforced a ban for placing political ads during the week before the Elections \cite{kovach_2020}, and indeed we did not locate any new ads in our collected sample for the specific period.  

To detect ads Facebook moderated, we crawled the ads in our dataset to locate on which Facebook placed a flag that specific that they were removed. In total, we located the removal of 8635 ads from 253 advertisers on the platform. 

For the ads in our collection that were in the form of image or video, we transformed them to text using the Google Cloud Vision and Speech-to-Text APIs, to make them available for further statistical processing. Next, we queried the \textit{Federal Election Commission} (FEC) database \cite{fec}, to investigate how many advertisers were registered as political entities. We also matched advertisers with their corresponding records in the political tracking website FollowTheMoney \cite{opensecrets_2022}, which classify them as Political Action Committees (PACs), Authorized Campaign Committees, NGOs, State related entities, Corporation or Labor entities, or other entities. Based on the information of the platform, we also categorized advertisers in respect to the content they created, i.e. whether they promoted a specific ideology or single issue, whether they were promoting civil rights, they were general advertising agencies, they created policy related content, they created candidate or party related content, they were selling merchandise, promoting the issues of government and state agencies, or they were asking individuals to perform civic service (e.g. working in election administration).

\subsection{Google \& YouTube}

Google's ad archive contains information about political ads placed through the Google ad services. This includes ads that appeared in the google search engine, on third party websites that use the services, and on YouTube. The archive provides information about the content of an ad, a lower and upper number of generated impressions and cost, as well as which user groups in terms of age, gender, and location were \textbf{targeted} by the advertiser. Like with Facebook, we crawled the ad archive and collected all political ads placed in the 2 months prior to the election (September 1st to November 4th) by advertisers who spent again at least hundred thousand dollars. The final dataset contains 117,607 ads from 490 advertisers, which represents 83\% of political ads placed during this period.

Unlike Facebook, Google removes ads that violate its terms \& conditions, leaving only its meta-data visible in the archive. To locate the content of ads that were removed, we systematically crawled the archive between September and November 2020, and once in January 2021. By comparing the crawls we located 9735 Ads that were moderated by Google. Similarly, we located 451 Ads that were also moderated, but we were not able to uncover their exact content, since they were removed prior to us collecting their original content.  We transformed all ads that were in the form of images and all YouTube videos into text using the Google Cloud Vision and Speech-to-Text APIs, to make them available for further statistical processing.  For the advertisers present in the dataset, we followed the same classification process as for Facebook, searching for their presence in the FEC database, and coding their type and advertising content. 

\subsection{TikTok} \label{tiktok_methods}

Formally, TikTok does not allow the placement of political ads. But we observed influencers engaged in political campaigning, who formed so-called HypeHouses. HypeHouses are TikTok accounts managed by coalitions of political influencers, generating content supporting specific candidates. 

We started with a list of known influencers and HypeHouses \cite{lorenz_2020} and by snowballing we collected other popular accounts that interacted with them. This resulted in a final list of 40 Democratic and 56 Republican HypeHouses and political influencers. We then crawled the HypeHouse videos between September 1st and October 15th. We had wanted to collect data through November 4th, but our access to crawl the platform was restricted, as the platform changed its internal API structure. (Appendix, table \ref{influencers}). Next, we created a list of political hashtags (Appendix, table \ref{queryt}) that included candidates’ names, election related issues such as mail-in ballots, and general political issues such as abortion or gun laws. We searched for videos containing these hashtags. Because our TikTok crawl returns only trending content and not all videos related to a hashtag, we identified video creators of the returned content and collected all of their videos for the same period as above. Our final dataset contained 2,690,923 videos from more than 61,000 TikTok creators. For each creator, we obtained information such as their account description, number of followers, and general popularity. For each video, we collected information about how many times they were viewed, liked, and shared, as well as their description. For the HypeHouses, we reviewed the profiles to manually categorize them based on whether they reported links to following entities: PACs, NGOs, politicians, media outlets and whether they were selling merchandise, or were asking for donations. For the purpose of evaluating its moderation practices, we relied on TikTok returning information about whether a video was an ad or was assigned a flag, such as being related to the US elections.

\section{Figures}

\begin{figure}[!htb]
    \centering
     \centering
    \subfloat{{\includegraphics[width=8cm]{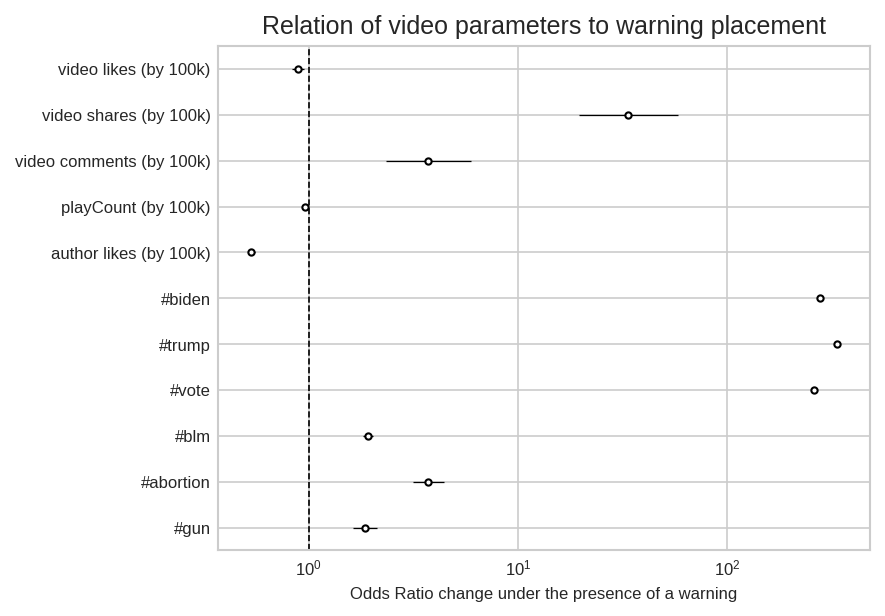} }}%
    \qquad
    \subfloat{\includegraphics[width=7cm]{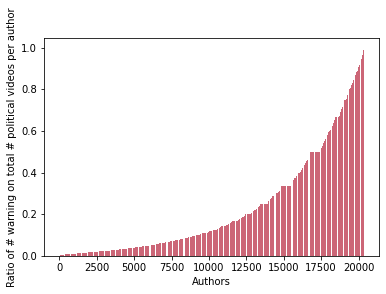}}
    \caption{Left: Forest plot of logistic regression model predicting whether a political TikTok video will be labeled with a warning flag. Right: Ratio of political videos flagged by user on TikTok.}
    \label{tiktok_moderated}
\end{figure}

\begin{figure*}[!htb]
    \centering
    \includegraphics[width=1\textwidth]{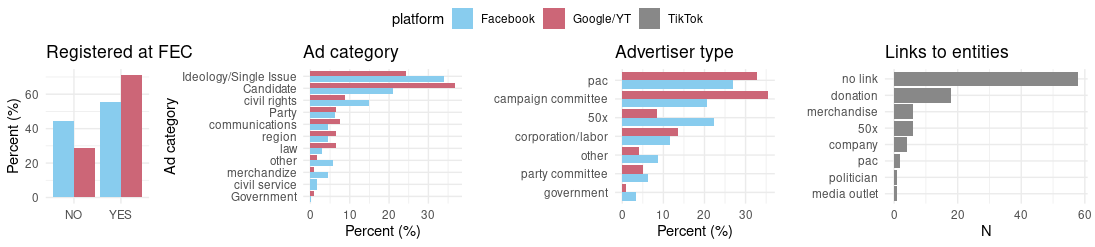}
    \caption{ Overview of visible advertisers in the libraries by platform in the dataset. Facebook is depicted in blue, Google is depicted in burgundy, and TikTok is depicted in grey. The bottom-right bar plot depicts the number of influencers linked to political entities.}
    \label{FEC}
\end{figure*}


\begin{table}[!htb]
\caption{We collected generated content from the following 96 Republican \& Democratic influencers. }
\begin{tabular}{cc} 
\multicolumn{2}{c}{ Influencers}      \\ \hline
Republican    & Democratic              \\ \hline \hline
\begin{tabular}[c]{@{}c@{}}nickvideos,thecjpearson,eliciawho,dylan.odin,redboyhickibilly,\\ machooch,theconservativevalues,tophertownmusic,samditzhazy,\\ bodittle,jsinnmusic,thebiasone,mommy\_nikki,rheannonfae,\\ conservativeHypeHouse,o\_rileyyyautoparts,kp.thepatriot,\\ c.j.\_production,therepublicanHypeHouse,donaldtrumpteam,\\ republicanism,matt4186,therealbentaylor,albertojdejesus,\\ youngrepub,thescoop\_us,jimjrpavv,thebadopinion,realjohndennis,\\ yourcity,c.jennings7152822,patriotfacts,dylanmaddentv,\\ frankynofingers,lamot11,kindall.k,gcnow,truthseeker5536,\\ daddy\_no\_pc,daddy\_no\_pc2,megaamerican,zc\_55,americanblondie,\\ thesavvytruth,therightlefty,christianwalk1r,matty.merica,\\ claytonkeirns,emmanuelharouno,the.rickytaylor,chadvideos,\\ therepublicangirlls,imtriggered,mattconvard,bobs\_politics,\\ youngrepublicans45,\end{tabular} & \begin{tabular}[c]{@{}c@{}}thecadelewis,chabella40,belessstupid,the.ghost88,imnotnatalie,\\ virtualconnectors,professorross,save.america,democrat\_me,\\ mr.shaw7,docd12,electro\_high,donthecon\_and\_associates,wadeslade,\\ thedemHypeHouse,thatliberalgirl,leftistjayce,futurestatesmanalexander,\\ somepoliticaldude,theleewithnoname,heathergtv,shashaankvideos,\\ thehumanrightsgroup,maya2960,theleftistdude,kaivanshroff,\\ maxwellblaine,typical\_democrat,j0emorris,izuhhhhhbel,thealanvargas,\\ spiicyboi7,lord\_timothais,deerodx,bidenssunglasses,\\ theprogressivepolicy,jbiii,liberalcorner,yaboihatestiktok,bidencoalition,\end{tabular} \\                      \hline
\end{tabular}
\label{influencers}
\end{table}

\begin{table}[!htb]
\caption{Regexp expression by which we matched TikTok videos' hashtags. }
\begin{tabular}{c}  Hashtags \\ \hline \begin{tabular}[c]{@{}l@{}}  \texttt{trump|biden|harris|fakenews|election|debate|maga|democrat|}\\ \texttt{republican|gun|libert|lgbt|conservative|politic|president|left|right|}\\ \texttt{vote|ballot|equality|kamala|bluewave|envelope|blm|blacklives|}\\ \texttt{alllives|dems|reps|settlefor|kag|alm} \texttt{|floyd|breonna|abortion|vax|}\\ \texttt{vaccine|factcheck|fakenews|aoc}  \end{tabular}  \\                \hline
\end{tabular}
\label{queryt}
\end{table}

\begin{table}[!htb]
\caption{Top advertisers on Facebook and Google who are not registered at the Federal Election Commission (FEC).}
\resizebox{12cm}{!}{\begin{tabular}{|ll|ll|}
\multicolumn{4}{c}{\textbf{Top Non FEC registered advertisers}}                                                                                                                                                                                                \\ \hline
\textbf{Google}                                                                                                                                                     & \textbf{Spent(\$)} & \textbf{Facebook}                      & \textbf{Spent(\$)} \\  \hline
YES ON 22 - SAVE APP-BASED JOBS \& SERVICES                                                                                                                                 & 7,122,400          & Yes on Prop 22                         & 5,382,534          \\
CONSERVATIVE BUZZ LLC                                                                                                                                                       & 5,849,400          & WhatsApp                               & 5,000,000          \\
SAHAK NALBANDYAN                                                                                                                                                            & 3,254,100          & No On Proposition 23                   & 4,302,798          \\
\begin{tabular}[c]{@{}l@{}}NO ON 23 - STOP THE DANGEROUS \& COSTLY \\ DIALYSIS PROPOSITION, A COALITION OF DIALYSIS\\  PROVIDERS, NURSES, DOCTORS AND PATIENTS\end{tabular} & 3,221,900          & When We All Vote                       & 4,014,869          \\
THERESA GREENFIELD FOR IOWA, INC.                                                                                                                                           & 3,193,900          & U.S. Census Bureau                     & 3,658,788          \\
NEWSMAX MEDIA INC                                                                                                                                                           & 2,716,300          & Democratic Governors Association (DGA) & 2,884,356          \\
EPOCH USA INC.                                                                                                                                                              & 2,154,300          & Voto Latino                            & 2,670,929          \\
ALLIANCE FOR A BETTER MINNESOTA ACTION FUND                                                                                                                                 & 1,229,800          & Stop the Illinois Tax Hike Amendment   & 2,378,970          \\
\begin{tabular}[c]{@{}l@{}}COALITION TO STOP THE PROPOSED\\  TAX HIKE AMENDMENT\end{tabular}                                                                                & 1,181,900          & The Collective PAC                     & 2,121,249          \\
JUDICIAL WATCH INC                                                                                                                                                          & 1,140,700          & One North Carolina                     & 2,033,109         \\ \hline
\end{tabular}}
\end{table}

\begin{table}
\caption{Removed ads in our dataset for each platform, by how many advertisers and how many impressions they generated prior to their removal.}
\label{removal_ads}
\resizebox{0.37\textwidth}{!}{\begin{tabular}{r|ccc}
\textbf{Platform} & \textbf{\% removed} & \textbf{reach}                                                  & \textbf{Advertisers} \\ \hline
\textbf{Google}   & 13.3\%              & \begin{tabular}[c]{@{}c@{}}150 mil. - \\ 1 billion\end{tabular} & 256 (18\%)           \\
\textbf{YouTube}  & 4.5\%               & \begin{tabular}[c]{@{}c@{}}120 mil. - \\ 1 billion\end{tabular} & 307 (22\%)           \\
\textbf{Facebook} & 1.2\%               & 200 million                                                     & 253 (30\%)          
\end{tabular}}
\end{table}


\begin{table}
\caption{Warnings placed on TikTok videos with at least one political hashtag.}
\begin{tabular}{ll}
\textbf{Label}                                           & \textbf{Counts} \\ \hline
Get info on the U.S. elections                           & 243,440         \\
Learn the facts about COVID-19                           & 2,341           \\
The action in this video could result in serious injury. & 30             
\end{tabular}
\end{table}

\begin{table}[!htb]
    \begin{minipage}[t]{.4\linewidth}
      \caption{Ordinal linear regression results for predicting the generated number of impressions for each advertisement of Biden \& Trump on Google. }
                    \label{regression1}

\resizebox{5cm}{!}{\begin{tabular}{c c c}
\hline
Variable & Estimator & St. Error \\
\hline
 \$100-\$1k           & $5.23^{***}$  
                      & $(0.05)$      \\
\$1k-\$50k           & $9.71^{***}$  
                      & $(0.07)$      \\
 \$50k-\$100k       & $15.37^{***}$ 
                      & $(0.22)$      \\
$>$\$100k          & $17.97^{***}$ 
                      & $(0.25)$      \\
Google Network         & $9.06^{***}$  
                      & $(0.09)$      \\
YouTube         & $5.22^{***}$  
                      & $(0.07)$      \\
Male                  & $-0.01$      
                      & $(0.22)$      \\
Female                & $0.08$        
                      & $(0.22)$      \\
Age 18-24           & $2.04^{***}$  
                      & $(0.29)$      \\
Age 25-34           & $-1.07^{***}$ 
                      & $(0.22)$      \\
Age 45-54           & $-0.88^{***}$ 
                      & $(0.13)$      \\
Age 55+           & $-0.39^{**}$ 
                      & $(0.14)$      \\
Zip code               & $-0.50^{***}$ 
                      & $(0.03)$      \\
County                & $-0.13^{*}$   
                      & $(0.06)$      \\
USA                   & $0.02$        
                      & $(0.06)$      \\
region not targeted & $-0.18$       
                      & $(0.13)$      \\
Trump over Biden                 & $0.34^{***}$  
                      & $(0.03)$      \\
$\leq$ 10k|10k-100k        & $9.41^{***}$  
                      & $(0.09)$      \\
10k-100k|100k-1M      & $14.30^{***}$ 
                      & $(0.12)$      \\
100k-1M|1M-10M        & $18.14^{***}$ 
                      & $(0.13)$      \\
1M-10M|$>$ 10M        & $22.26^{***}$ 
                      & $(0.23)$      \\
\hline
AIC                   & $50018.68$  &  \\
BIC                   & $50208.89$  &  \\
Log Likelihood        & $-24988.34$  & \\
Deviance              & $49976.68$   & \\
Num. obs.             & $63422$      & \\
\hline
\multicolumn{2}{l}{\scriptsize{$^{***}p<0.001$; $^{**}p<0.01$; $^{*}p<0.05$}}
\end{tabular}}
    \end{minipage}%
    \quad
    \begin{minipage}[t]{.5\linewidth}
      \centering
        \caption{Linear regression results predicting the impression/cost ratio for ads placed by Biden \& Trump on Facebook.}
            \label{regression2}

\resizebox{10cm}{!}{\begin{tabular}{cccccccc}
\hline
 Variable & Estimator& &St. Error & Variable & Estimator& &St. Error \\
\hline
AL                        & $7.53^{*}$     &
                          & $(3.48)$       &
AK                        & $4.13$         &
                          & $(6.31)$       \\
AZ                        & $1.22$         &
                          & $(2.80)$       &
AR                        & $16.94^{***}$  &
                          & $(4.40)$       \\
CA                        & $-1.10$        &
                          & $(2.87)$       &
CO                        & $8.56^{**}$    &
                          & $(2.89)$       \\
CT                        & $-3.28$        &
                          & $(4.27)$       &
DE                        & $-0.98$        &
                          & $(6.68)$       \\
FL                        & $3.43$         &
                          & $(2.80)$       &
GA                        & $2.99$         &
                          & $(2.80)$       \\
ID                        & $24.48^{***}$  &
                          & $(5.43)$       &
IL                        & $3.50$         &
                          & $(3.19)$       \\
IN                        & $9.19^{**}$    &
                          & $(3.37)$       &
IA                        & $2.19$         &
                          & $(2.81)$       \\
KS                        & $22.58^{***}$  &
                          & $(4.81)$       &
KY                        & $30.35^{***}$  &
                          & $(4.12)$       \\
LA                        & $6.57$         &
                          & $(3.42)$       &
ME                        & $-0.47$        &
                          & $(2.83)$       \\
MD                        & $1.19$         &
                          & $(3.43)$       &
MA                        & $-13.80^{***}$ &
                          & $(3.44)$       \\
MI                        & $3.41$         &
                          & $(2.80)$       &
MN                        & $0.93$         &
                          & $(2.81)$       \\
MS                        & $22.91^{***}$  &
                          & $(4.54)$       &
MO                        & $19.84^{***}$  &
                          & $(3.80)$       \\
MT                        & $-1.43$        &
                          & $(4.95)$       &
NE                        & $-2.11$        &
                          & $(2.83)$       \\
NV                        & $-0.72$        &
                          & $(2.81)$       &
NH                        & $-0.30$        &
                          & $(2.89)$       \\
NJ                        & $0.49$         &
                          & $(3.54)$       &
NM                        & $-0.48$        &
                          & $(5.64)$       \\
NY                        & $-8.92^{**}$   &
                          & $(3.09)$       &
NC                        & $1.89$         &
                          & $(2.80)$       \\
ND                        & $21.78^{***}$  &
                          & $(6.56)$       &
OH                        & $4.09$         &
                          & $(2.81)$       \\
OK                        & $17.90^{***}$  &
                          & $(4.17)$       &
OR                        & $0.31$         &
                          & $(3.40)$       \\
PA                        & $1.36$         &
                          & $(2.80)$       &
RI                        & $-10.22$       &
                          & $(6.91)$       \\
SC                        & $17.96^{***}$  &
                          & $(3.98)$       &
SD                        & $1.26$         &
                          & $(3.63)$       \\
TN                        & $19.89^{***}$  &
                          & $(3.64)$       &
TX                        & $5.18$         &
                          & $(2.85)$       \\
UT                        & $5.22$         &
                          & $(3.46)$       &
VT                        & $-3.84$        &
                          & $(3.84)$       \\
VA                        & $8.27^{**}$    &
                          & $(2.86)$       &
WA                        & $-3.66$        &
                          & $(3.25)$       \\
WV                        & $11.07^{*}$    &
                          & $(5.19)$       &
WI                        & $0.65$         &
                          & $(2.80)$       \\
WY                        & $11.90$        &
                          & $(6.51)$       &
DC                        & $-26.14^{***}$ &
                          & $(5.52)$       \\
Male 18-24                   & $2.06$         &
                          & $(1.97)$       &
Male 25-34                   & $7.70^{***}$   &
                          & $(1.90)$       \\
Male 35-44                   & $7.88^{***}$   &
                          & $(1.91)$       &
Male 45-54                   & $7.05^{***}$   &
                          & $(1.91)$       \\
Male 55-64                   & $2.35$         &
                          & $(1.91)$       &
Male 65+                     & $-5.67^{**}$   &
                          & $(1.92)$       \\
Female 18-24                   & $1.43$         &
                          & $(1.96)$       &
Female 25-34                   & $7.59^{***}$   &
                          & $(1.91)$       \\
Female 35-44                   & $8.95^{***}$   &
                          & $(1.92)$       &
Female 45-54                   & $6.49^{***}$   &
                          & $(1.93)$       \\
Female 55-64                   & $0.71$         &
                          & $(1.91)$       &
Female 65+                     & $-5.37^{**}$   &
                          & $(1.90)$       \\
Ad delivery start time & $-0.17^{***}$  &
                          & $(0.00)$       &
Biden Campaign            & $27.53^{***}$  &
                          & $(3.35)$       \\
Trump Campaign            & $18.27^{***}$  &
                          & $(3.35)$       \\
\hline
\multicolumn{2}{l}{\scriptsize{$^{***}p<0.001$; $^{**}p<0.01$; $^{*}p<0.05$}}
\end{tabular}}
    \end{minipage} 

\end{table}

\begin{table}[!htb]
    \centering
    
    \caption{Logistic regression results for predicting whether a TikTok video contains an election related warning.}

\begin{center}
\resizebox{9cm}{!}{\begin{tabular}{rcccccc}
               & \textbf{coef} & \textbf{std err} & \textbf{z} & \textbf{P$> |$z$|$} & \textbf{[0.025} & \textbf{0.975]}  \\
\midrule
\textbf{video likes (by 100k)}    &      -0.1177  &        0.031     &    -3.738  &         0.000        &       -0.179    &       -0.056     \\
\textbf{video shares (by 100k)}    &       3.5197  &        0.278     &    12.669  &         0.000        &        2.975    &        4.064     \\
\textbf{video comments (by 100k)}    &       1.3187  &        0.237     &     5.566  &         0.000        &        0.854    &        1.783     \\
\textbf{playCount (by 100k)}    &      -0.0431  &        0.006     &    -7.270  &         0.000        &       -0.055    &       -0.032     \\
\textbf{author likes (by 100k)}    &      -0.6334  &        0.019     &   -33.465  &         0.000        &       -0.671    &       -0.596     \\
\textbf{\#biden}    &       5.6269  &        0.010     &   555.612  &         0.000        &        5.607    &        5.647     \\
\textbf{\#trump}    &       5.8192  &        0.018     &   319.928  &         0.000        &        5.784    &        5.855     \\
\textbf{\#vote}    &       5.5635  &        0.018     &   312.633  &         0.000        &        5.529    &        5.598     \\
\textbf{\#blm}    &       0.6550  &        0.028     &    23.246  &         0.000        &        0.600    &        0.710     \\
\textbf{\#abortion}   &       1.3162  &        0.087     &    15.117  &         0.000        &        1.146    &        1.487     \\
\textbf{\#gun}   &       0.6233  &        0.068     &     9.172  &         0.000        &        0.490    &        0.756     \\
\textbf{const} &      -4.5996  &        0.008     &  -570.047  &         0.000        &       -4.615    &       -4.584     \\
\bottomrule
\end{tabular}}
\end{center}
    \label{tab:my_label}
\end{table}

\end{document}